\documentclass[journal]{IEEEtran}

\usepackage{xcolor,soul,framed} 
\colorlet{shadecolor}{yellow}
\usepackage{graphicx}
\graphicspath{{figures/}{../pdf/}{../jpeg/}}
\DeclareGraphicsExtensions{.pdf,.jpeg,.png}
\usepackage[cmex10]{amsmath}
\usepackage{latexsym,bm,amssymb}
\usepackage{array}
\usepackage{mdwmath}
\usepackage{mdwtab}
\usepackage{eqparbox}
\usepackage{url}
\usepackage{enumerate}
\usepackage{amsfonts}
\usepackage{algorithmic}
\usepackage{algorithm}
\usepackage[numbers,sort&compress]{natbib}
\usepackage[justification=centering]{caption}
\usepackage{mathrsfs}
\usepackage{diagbox}

\usepackage{multirow}

\usepackage{booktabs} 

\usepackage{footnote}

\usepackage{nomencl}
\makenomenclature
\usepackage{ifthen}
  \renewcommand{\nomgroup}[1]{%
  \item[\bfseries
  \ifthenelse{\equal{#1}{A}}{Symbols of UC}{%
  \ifthenelse{\equal{#1}{S}}{Symbols of CL}{}}%
  ]}

\renewcommand{\baselinestretch}{1}

\usepackage{tikz}
\usetikzlibrary{positioning,arrows.meta,shapes.geometric,fit,calc}

\begin{document}

\captionsetup{font={small}}

\title{Voltage Stability Assessment with Path-Coupled Load Growth and Corrective Generator Response}

\author{Lanqing Shan,~\IEEEmembership{Member,~IEEE,}
\thanks{This work was supported in part by National Key R\&D Program of China (No. X) and Technical Project of the State Grid: (Corresponding author: X.)}

}

\markboth{Submitted to IEEE Trans. Power Systems}%
{Shell \MakeLowercase{\textit{et al.}}: Bare Demo of IEEEtran.cls for IEEE Journals}

\maketitle

\begin{abstract}
	Voltage stability margin assessment is essential for the secure operation of renewable-dominated power systems. Conventional continuation-based methods evaluate the margin along predefined load-growth paths with fixed generator participation, while practical operation allows generators to be redispatched to alleviate voltage stress and reshape the power-flow trajectory as the system approaches voltage collapse. This paper proposes a path-coupled margin assessment approach that incorporates corrective generator response into static voltage stability margin assessment. In the proposed approach, the load-growth direction and generator response direction are simultaneously determined at each continuation step, enabling the assessment trajectory to account for generator response while tracing the system toward voltage collapse. The voltage stability margin is then evaluated by the cumulative active load increase along this coupled trajectory. Based on the obtained trajectory and collapse point, a feasible redispatch direction is further derived to improve the margin of the current operating state. The economic cost of voltage stability enhancement is quantified through a marginal stability cost, providing an economic indicator for additional stability support. Case studies on various test systems demonstrate that the proposed framework can effectively capture the impact of corrective generator redispatch on voltage stability assessment, provide effective guidance for margin enhancement, and quantify the cost associated with voltage stability improvement.
\end{abstract}


\IEEEpeerreviewmaketitle

\section{Introduction}\label{introduction}

\IEEEPARstart{P}{ower} systems with high renewable and inverter-based generation are increasingly exposed to voltage stability risks \cite{hatziargyriou2021definition}. It is becoming increasingly important to evaluate how much additional loading can be accommodated before the system reaches a voltage collapse point \cite{benato2022effect, chintakindi2024wide, yang2025closest}. This quantity, commonly referred to as the voltage stability margin, is inherently path-dependent because different load-growth and power-injection variation paths may lead to different collapse points \cite{sode2006maximum, neves2020computation}. Conventional continuation-based assessments usually compute this margin along a prescribed loading direction with a predetermined generator participation rule \cite{ajjarapu1992continuation}. Such a treatment provides an effective way to trace the power-flow trajectory, but the predefined path cannot fully represent the adaptive generation adjustments involved in practical operation. As the system approaches voltage collapse, generators can adjust active and reactive outputs to compensate for load increases and alleviate voltage stress, thereby shaping the continuation trajectory toward voltage collapse \cite{capitanescu2009coupling, chu2022voltage}. Therefore, incorporating corrective generator response into the margin assessment process is essential for obtaining a more realistic evaluation of the distance from the current operating point to voltage collapse.

Incorporating corrective generator response into static voltage stability assessment introduces additional coupling in the continuation calculation. At each operating point, a given load-growth pattern can be paired with different feasible generator responses, and their combination determines the resulting net injection direction. The feasible response is further restricted by active-power balance and generator active- and reactive-power limits, whose remaining regulation ranges evolve along the continuation trajectory \cite{jia2023voltage, wang2023voltage}. Therefore, the load-growth and generator-response directions need to be determined jointly and updated during continuation rather than prescribed independently before the assessment.

Existing work has progressively relaxed the path direction assumptions in static voltage stability margin assessment \cite{wu2021searching, yang2021monitoring,wu2024approximating}. Classical continuation power flow (CPF) methods compute the voltage collapse point along a prescribed loading path \cite{ajjarapu1992continuation, zimmerman2011matpower}. To better capture practical operating characteristics along the prescribed path, extended continuation methods incorporate renewable-generation characteristics into Jacobian formulations \cite{wang2022extended}, and smooth power-flow models represent operating limits and limit-induced bifurcations within a unified differentiable framework \cite{neves2022smooth}. Although these methods provide more realistic collapse-point calculations, the loading path remains specified before the assessment. To further reduce the dependence on predefined paths, closest-bifurcation and shortest-distance methods search for adverse load-growth directions toward the voltage-collapse boundary \cite{dobson1993new}. Some studies restrict the search space according to credible loading uncertainties or reactive power limits \cite{lin2020static, wang2024iterative}, while recent optimization-based methods further incorporate feasible operating constraints into the assessment problem \cite{wang2025assessing}. These approaches extend voltage stability assessment from prescribed-path evaluation toward uncertainty-aware and feasibility-constrained direction search \cite{alzubaidi2022impact}. However, the generator response along the variation path is still determined by predefined balancing rules or participation factor.

Generation adjustments have also been investigated as a means of improving voltage stability margins \cite{wang2000re, jia2024learning}. Sensitivity-based methods quantify the impact of generation adjustments on stability margins and use this information to guide preventive active- and reactive-power redispatch \cite{greene1997sensitivity,capitanescu2002preventive}. Based on these sensitivity evaluations, voltage-stability-constrained optimal power flow formulations further incorporate stability requirements into dispatch optimization \cite{cui2018new, wang2018sdp}. Recent studies have investigated the co-optimization of generation dispatch and PV-PQ bus type profiles under Jacobian-based stability constraints \cite{song2023voltage}, as well as the integration of CPF-based load-margin constraints into dispatch decisions \cite{liu2024multi, chevalier2026identifying}. These methods improve voltage stability by optimizing generator adjustments with respect to an established stability metric or loading condition. However, they generally determine generator response after the loading trajectory has been specified, without explicitly considering how to determine generator response to alleviate the increaseing voltage stress along the path-coupled trajectory between load growth and generator response.

To address the above gap, this paper proposes a path-coupled voltage stability margin assessment method that incorporates corrective generator response into static voltage stability margin assessment. The proposed method determines generator active and reactive power adjustments as part of the continuation trajectory. At each continuation step, the adverse load-growth direction and corrective generator response direction are jointly determined under system operating constraints, and the coupled trajectory is traced until voltage collapse. The path-coupled voltage stability margin is evaluated by the accumulated active load increase along this trajectory. Based on this margin assessment, a redispatch-based margin improvement method is further developed for the current operating state, and the associated economic cost is quantified through a marginal stability cost. The main contributions are summarized as follows:
\begin{enumerate}
\item A path-coupled margin assessment method is proposed for static voltage stability. The method jointly determines the load-growth direction and geneartor response direction along the continuation trajectory and evaluates the margin as the accumulated active load increase from the current operating state to voltage collapse.
\item A margin-improving redispatch direction is derived for voltage stability enhancement at the current operating state. The direction determines generator active and reactive power adjustments that increase the path-coupled voltage stability margin while satisfying operating constraints.
\item A marginal stability cost evaluation is proposed to quantify the economic cost of voltage stability enhancement. By relating redispatch cost changes to the resulting margin improvement, the proposed metric provides an economic indicator for additional voltage stability support.
\end{enumerate}

The remainder of this paper is organized as follows. Section \ref{framework} presents the proposed path-coupled voltage stability assessment framework. Section \ref{sec_c1} illustrates the path-coupled margin assessment method considering adverse load growth and corrective generator response. Section \ref{sec_c2_c3} gives the redispatch-based margin improvement method and the marginal stability-cost evaluation. Section \ref{sec_case} performs case studies on various test systems. Section \ref{sec_conclusion} concludes the paper.

\section{Path-Coupled Assessment Framework}\label{framework}

A margin assessment considering corrective generator response requires the continuation path to be modeled as a decision-dependent trajectory. To this end, the proposed framework formulates a coupled process in which adverse load growth and corrective generator response are jointly determined at each continuation step to construct the continuation trajectory toward voltage collapse.

\subsection{Path-Coupled Margin Assessment Problem}

Static voltage stability is analyzed with the steady-state power flow equations. For bus \(i\), the active and reactive power balance equations can be written as
\begin{equation}
	\begin{aligned}
		P_i
		-
		V_i\sum_{j\in\mathcal{N}}
		V_j\left(
		G_{ij}\cos\theta_{ij}
		+
		B_{ij}\sin\theta_{ij}
		\right)
		&=0,\\
		Q_i
		-
		V_i\sum_{j\in\mathcal{N}}
		V_j\left(
		G_{ij}\sin\theta_{ij}
		-
		B_{ij}\cos\theta_{ij}
		\right)
		&=0,
	\end{aligned}
	\label{eq:pf_nodal}
\end{equation}
where \(\theta_{ij}=\theta_i-\theta_j\), and \(G_{ij}\) and \(B_{ij}\) are the real and imaginary parts of the bus admittance matrix. Stacking the nodal equations in Eq. \eqref{eq:pf_nodal} gives \(f(x;\rho)=0\), where \(x=[\theta,V]\) denotes the power-flow state and \(\rho=[P,Q]\) denotes the vector of nodal active and reactive power injections. The static voltage stability boundary \(\mathcal{S}\) is the set of injection vectors at which the corresponding power-flow Jacobian becomes singular. Conventional continuation assessment evaluates the distance from the current operating point \(\rho^{(0)}\) to \(\mathcal{S}\) along a prescribed loading path, so the assessed margin depends on the loading pattern and generator participation rule specified before the continuation starts.

To describe corrective generator response during loading, this paper decomposes the local continuation direction in the injection space into a load-growth component \(d_L\) and a generator-response component \(d_G\). The component \(d_L\) describes active and reactive load increase, and is selected from admissible load-growth patterns to represent the loading variation that most severely stresses voltage stability. The component \(d_G\) describes the active and reactive generation adjustment paired with this loading stress, mitigating voltage-stability deterioration through power-flow redistribution and voltage support. At the \(k\)-th continuation step, the two components are determined at the current operating point and jointly update the injection vector through Eq. \eqref{eq_update}, where \(\Delta\lambda\) is the continuation step size. Repeating this process generates a path-coupled continuation trajectory with a locally updated direction at each operating point.
\begin{equation}
	\rho^{(k+1)}
	=
	\rho^{(k)}
	+
	\Delta\lambda
	\left(d_L^{(k)}+d_G^{(k)}\right) \label{eq_update}
\end{equation}

The path-coupled voltage stability margin is defined as the cumulative active load increase along the generated trajectory before it reaches \(\mathcal{S}\). Under the generation convention, active load growth appears as a decrease in net active power injection. Hence, if \(d_{L,P}^{(k)}\) denotes the active-power part of \(d_L^{(k)}\), the margin is given by:
\begin{equation}
	\eta_{\mathrm{pc}}(\rho^{(0)})
	=
	\sum_{k}
	\left(
	-\mathbf{1}^{\top}\Delta\lambda d_{L,P}^{(k)}
	\right),
\end{equation}

With this definition, generator response is no longer merely a prescribed participation rule in a continuation study. It becomes an endogenous component of the margin assessment, evaluated together with adverse load growth as the system moves toward voltage collapse. The resulting trajectory reflects how loading stress and generation adjustment jointly shape the approach to voltage collapse. This information provides a basis for improving the current operating point through generator response, while the associated change in operating cost makes it possible to quantify the cost of additional voltage stability gained through this adjustment.

\subsection{Overall Framework}

\begin{figure}[!t]
	\centering
	\includegraphics[width=\columnwidth]{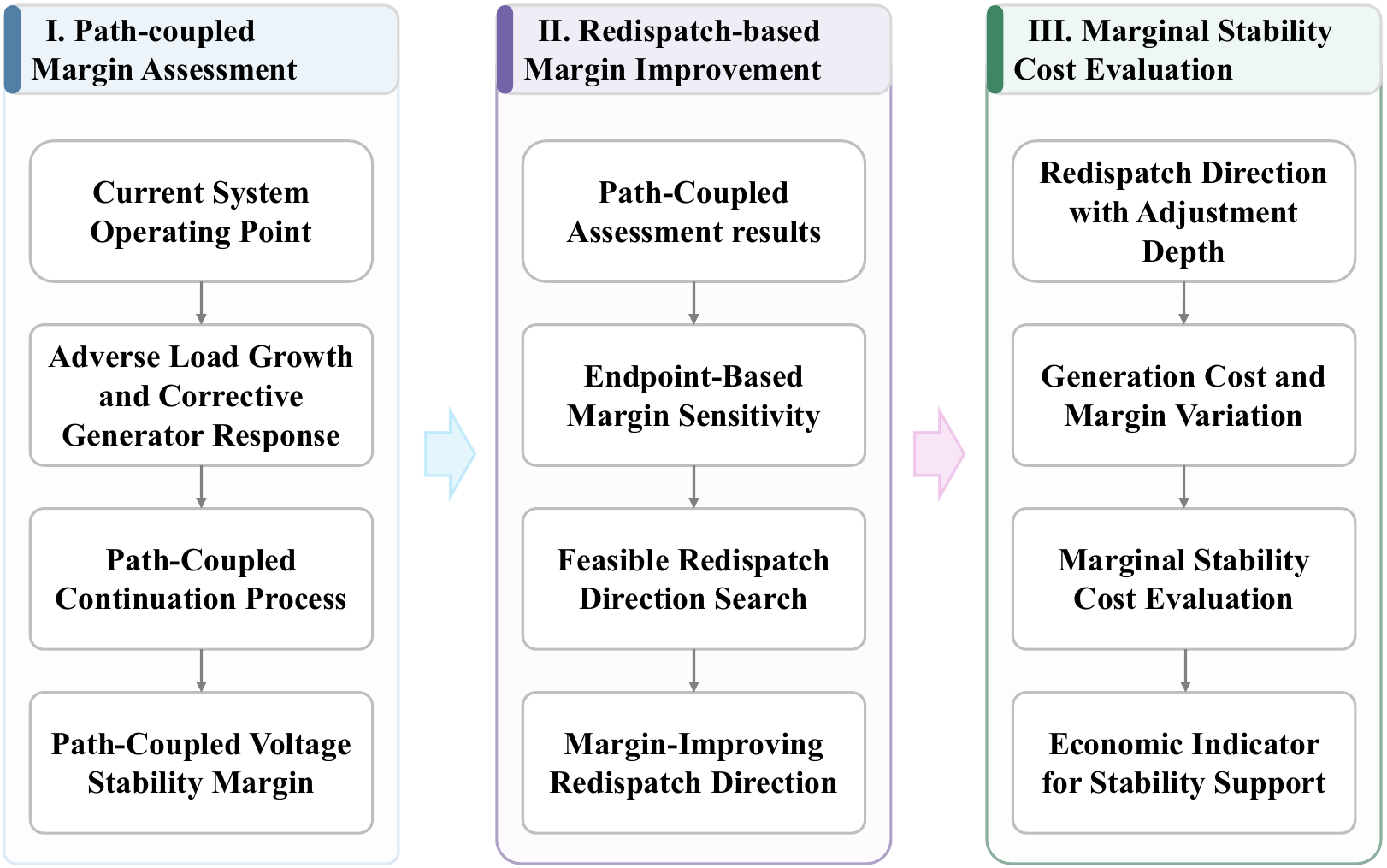}
	\caption{Overall framework of the proposed path-coupled voltage stability assessment with redispatch-based margin improvement and stability-cost evaluation.}
	\label{fig_framework}
\end{figure}

To achieve these objectives, this paper develops a three-stage framework, as illustrated in Fig.~\ref{fig_framework}.

The first stage builds the continuation trajectory. In the built trajectory, each operating point formulates a joint direction-selection problem over admissible load-growth patterns and feasible generator response. The load-growth component identifies the adverse loading direction, while generator active- and reactive-power adjustments counteract the resulting voltage-stability deterioration. The resulting net injection direction is embedded in a predictor-corrector continuation step. Repeating the direction selection and continuation update until voltage collapse yields the path-coupled margin as the cumulative active-load increase. The detailed formulation is given in Section~\ref{sec_c1}.

The second stage uses the path-coupled assessment results to derive a margin-improving redispatch direction for the current operating point. Since the assessed margin depends on the entire continuation trajectory, directly evaluating its variation with respect to generator adjustments is not straightforward. To solve this problem, the collapse point reached along the assessed trajectory, together with the corresponding margin, is used to construct an endpoint-based approximation of the margin sensitivity with respect to generator active and reactive outputs. This sensitivity is then used to determine a feasible redispatch direction subject to generator capability limits and active-power balance requirements. The resulting direction provides a feasible preventive redispatch for increasing the path-coupled margin at the current operating point. The detailed derivation is given in Section~\ref{sub_sec_c2}.

The third stage evaluates the economic cost associated with the margin improvement obtained from Stage II. Along the derived redispatch direction, the generator cost function provides the corresponding operating-cost variation, while the margin sensitivity obtained in Stage II quantifies the resulting voltage stability margin improvement. Their ratio defines the marginal stability cost at the current operating point. This metric measures the operating-cost increase required for one additional unit of voltage stability margin and provides an economic indicator for corrective stability improvement. The detailed definition and calculation are presented in Section~\ref{sub_sec_c3}.

\section{Path-Coupled Margin Assessment}\label{sec_c1}

The computation of the path-coupled margin centers on how the continuation direction is generated at each operating point. Different load-growth patterns may stress the system differently, while feasible generator response can relieve part of this stress through active- and reactive-power adjustments. To jointly account for adverse load growth and corrective generator response, this section develops a direction-selection approach that determines the load-growth and generator-response components of the net injection direction at each operating point along the continuation trajectory. The selected direction then drives the path-coupled continuation and determines the assessed voltage stability margin.

\subsection{Local Voltage Stability Sensitivity to Injection Directions}\label{sec_c1_setup}

At each operating point along the continuation trajectory, the next continuation direction needs to be selected according to how different feasible injection directions affect the proximity to voltage collapse. Recent studies have investigated the identification of critical injection directions within physically feasible operating regions \cite{wang2025assessing}. These approaches typically search for directions associated with the steepest nose-curve slopes to approximate the shortest distance to voltage collapse. However, a steep voltage response does not necessarily indicate the direction along which the power-flow equations approach loss of solvability most rapidly. Since voltage collapse is fundamentally associated with the singularity of the power-flow Jacobian, a more direct local criterion can be established by examining how a candidate injection direction affects the smallest singular value of the Jacobian. This criterion is developed below and subsequently used to select the adverse load growth and corrective generator response directions.

Consider the current operating point and let \(J=\partial \rho/\partial x\) denote the power-flow Jacobian. Its smallest singular value, denoted by \(\sigma=\sigma_{\min}(J)\), approaches zero as the system approaches a Jacobian singularity. For a candidate injection direction \(d\), the local injection path can be parameterized as \(\rho(\lambda)=\rho+\lambda d\). The first-order change in the smallest singular value along this direction can then be expressed as
\begin{equation}
	\frac{\partial \sigma}{\partial \lambda}
	=
	l^{\top}\frac{\partial J}{\partial \lambda}r
	=
	\sum_i
	l^{\top}
	\frac{\partial J}{\partial x_i}
	r\frac{\partial x_i}{\partial \lambda}
	=
	\mu^{\top}J^{-1}d
	=
	c^{\top}d ,
	\label{eq:sigma_variation}
\end{equation}
where \(l\) and \(r\) are the normalized left and right singular vectors associated with \(\sigma\), \(\mu_i=l^{\top}(\partial J/\partial x_i)r\), and \(c=J^{-\top}\mu\). Therefore, \(c^{\top}d\) directly characterizes the local effect of a candidate injection direction on the smallest singular value.

For a candidate continuation direction \(d\), the first-order change of the smallest singular value is therefore \(c^{\top}d\). If \(c^{\top}d<0\), the direction reduces \(\sigma\) and locally drives the system toward voltage collapse. If \(c^{\top}d>0\), the direction increases \(\sigma\) and provides local stability relief. Accordingly, the local stability degradation induced by \(d\) is defined as
\begin{equation}
	D(d)=-c^{\top}d .
	\label{eq:local_degradation}
\end{equation}
which provides the basis for determining the adverse load-growth component and the corrective generator response component of the continuation direction.

\subsection{Coupled Direction Optimization for Load Growth and Generator Response}

The local degradation rate in Eq. \eqref{eq:local_degradation} provides the criterion for choosing the continuation direction. For an adverse margin assessment, the loading component should represent the most severe stress while considering available generator support. At the \(k\)-th operating point, we utlize \(p_L\) to denote the active load-increase direction over load buses, while the reactive load increase is determined accordingly through the load power-factor relation. For generator response, the active and reactive directions are respectively denoted as \(g_P\) and \(g_Q\). Under the generation convention, the load increase and generator response are respectively represented by injection-space components \(d_L=-E_Lp_L\) and \(d_G=E_Pg_P+E_Qg_Q\), where \(E_L\), \(E_P\), and \(E_Q\) map load growth and generator response directions to the nodal injection vector. Substituting \(d_L+d_G\) into Eq. \eqref{eq:local_degradation} gives the core direction-selection model:
\begin{equation}
	\begin{aligned}
		\max_{p_L\in\mathcal{L}^{(k)}}\
		\min_{(g_P,g_Q)\in\mathcal{G}^{(k)}(p_L)}
		\quad
		 \alpha^{\top}p_L-\beta^{\top}g_P-\gamma^{\top}g_Q,\\
		\alpha=E_L^{\top}c,\quad
		\beta=E_P^{\top}c,\quad
		\gamma=E_Q^{\top}c .
	\end{aligned}
	\label{eq:direction_basic}
\end{equation}
where vector \(\alpha\) gives the local stability degradation effect of load growth at different buses, while \(\beta\) and \(\gamma\) give the local stability relief effect of generator active and reactive power response. The max-min formulation represents the coupling between adverse load growth and corrective generator response. For each candidate loading direction, the inner minimization determines its best feasible corrective generator response, while the outer maximization identifies the loading direction that is most adverse after accounting for this response.

To use the local max-min criterion in a continuation step, the direction variables are further restricted by the operating constraints at the current point. This gives the constrained direction-selection problem:
\begin{equation}
	\begin{aligned}
		\max_{p_L}\ \min_{g_P,g_Q}\quad
		&
		\alpha^{\top}p_L-\beta^{\top}g_P-\gamma^{\top}g_Q  \\
		\mathrm{s.t.}\quad
		&
		p_L\ge0,\quad
		\|p_L\|_2=1,\quad
		\mathbf{1}^{\top}p_L=\mathbf{1}^{\top}g_P,\\
		&
		\underline g_P^{(k)}\le g_P\le \overline g_P^{(k)},\quad
		\underline g_Q^{(k)}\le g_Q\le \overline g_Q^{(k)} .
	\end{aligned}
	\label{eq:direction_model}
\end{equation}
where \(p_L\ge0\) enforces nonnegative load increase, and \(\|p_L\|_2=1\) ensures a unit normalized loading direction. To ensure power balance, we use the equality \(\mathbf{1}^{\top}p_L=\mathbf{1}^{\top}g_P\) to balance the active load growth through generator active-power response. The response bounds in \eqref{eq:direction_model} are determined by the remaining active and reactive regulation ranges of the generators, e.g., \(\overline g_{P,i}^{(k)}=P_{G,i}^{\max}-P_{G,i}^{(k)}\) and \(\underline g_{P,i}^{(k)}=P_{G,i}^{\min}-P_{G,i}^{(k)}\), with analogous definitions for \(g_Q\).

The structure of Eq. \eqref{eq:direction_model} allows the max-min problem to be reduced to a scalar search, avoiding the direct solution of the coupled high-dimensional problem at each continuation step. We introduce \(b=\mathbf{1}^{\top}p_L=\mathbf{1}^{\top}g_P\) to represent the aggregate active load-growth amount in a candidate direction. For a fixed \(b\), the direction-selection problem separates into three allocation subproblems:
\begin{align}
	\phi_L(b)
	&=
	\max_{p_L\ge0,\ \|p_L\|_2=1,\ \mathbf{1}^{\top}p_L=b}
	\alpha^{\top}p_L,\label{eq:load_growth_b}\\
	\phi_P(b)
	&=
	\min_{\substack{
			\underline g_P^{(k)}\le g_P\le\overline g_P^{(k)}\\
			\mathbf{1}^{\top}g_P=b}}
	\left\{
	-\beta^{\top}g_P
	+
	\frac{\tau_P}{2}\|g_P-bw_P\|_2^2
	\right\},\label{eq:active_redispatch_b}\\
	\phi_Q(b)
	&=
	\min_{\underline g_Q^{(k)}\le g_Q\le\overline g_Q^{(k)}}
	\left\{
	-\gamma^{\top}g_Q
	+
	\frac{\tau_Q}{2}\|g_Q-bw_Q\|_2^2
	\right\}. \label{eq:reactive_redispatch_b}
\end{align}
where \(\tau_P>0\) and \(\tau_Q>0\) are regularization weights for the active- and reactive-power response patterns, respectively. Eq. \eqref{eq:load_growth_b} determines the allocation of load growth, while Eq. \eqref{eq:active_redispatch_b} and \eqref{eq:reactive_redispatch_b} determine the active- and reactive-power generator responses. The reference vectors in the quadratic terms are defined as
\begin{equation}
w_P=\frac{[\beta]_+}{\mathbf{1}^{\top}[\beta]_+},\qquad
w_Q=\kappa_Q\frac{[\gamma]_+}{\mathbf{1}^{\top}[\gamma]_+},
\label{eq:reference_pattern}
\end{equation}
where \(\kappa_Q\) denotes the prescribed aggregate reactive-to-active ratio associated with the load-growth power factor. These sensitivity-informed reference patterns in the quadratic terms help avoid overly concentrated generator responses while retaining the local tendency of increasing \(\sigma_{\min}\).

We utilize \(\Psi(b)=\phi_L(b)+\phi_P(b)+\phi_Q(b)\) denote the scalar value function obtained after optimizing the load-growth and generator response in Eq. \eqref{eq:load_growth_b}-\eqref{eq:reactive_redispatch_b}. This value function can be evaluated efficiently because the three allocation subproblems admit KKT-based normalization and projection solutions. The direction-selection problem is therefore reduced to choosing \(b\) through the one-dimensional search:
\begin{equation}
	b^{\star}\in\arg\max_{b\in\mathcal{B}^{(k)}}\Psi(b),
	\label{eq:b_search}
\end{equation}
where \(\mathcal{B}^{(k)}\) is the feasible interval of \(b\) for which the three allocation subproblems are all feasible. After \(b^{\star}\) is obtained, the corresponding optimizers \(p_L^{\star}(b^{\star})\), \(g_P^{\star}(b^{\star})\), and \(g_Q^{\star}(b^{\star})\) define \(d_L^{(k)}=-E_Lp_L^{\star}(b^{\star})\) and \(d_G^{(k)}=E_Pg_P^{\star}(b^{\star})+E_Qg_Q^{\star}(b^{\star})\). The net injection direction \(d_L^{(k)}+d_G^{(k)}\) is then used in the continuation step.

\subsection{Path-Coupled Continuation and Margin Evaluation}

Algorithm \ref{alg:pc_margin} summarizes the proposed path-coupled margin assessment (PCMA). Starting from the base-case solution, the algorithm repeatedly updates the Jacobian-based sensitivity coefficients, solves the direction-selection problem, and performs a continuation step along the adaptively selected net injection direction with a small step length \(\Delta\lambda\), i.e., \(\rho^{(k+1)}=\rho^{(k)}+\Delta\lambda d^{(k)}\). Since the direction-selection problem is resolved at each operating point, the adverse load-growth pattern and the corrective generator response are updated along the traced trajectory. At step \(k\), \(b_k^{\star}\) gives the aggregate active load-growth amount in the normalized direction, so \(\Delta\lambda b_k^{\star}\) represents the active load-growth increment contributed by this step. These increments are accumulated until the power-flow Jacobian approaches singularity, indicated by \(\sigma_{\min}(J^{(k)})\le \epsilon_{\sigma}\), and the accumulated value is reported as the path-coupled voltage stability margin.

\begin{algorithm}[t]
	\caption{Path-Coupled Margin Assessment}
	\label{alg:pc_margin}
	\begin{algorithmic}[1]
	\STATE \textbf{Input:} base-case solution \((x^{(0)},\rho^{(0)})\), generator limits, load power factors, step length \(\Delta\lambda\), and tolerance \(\epsilon_{\sigma}\).
	\STATE Set \(k=0\) and \(\eta_{\mathrm{pc}}=0\).
	\WHILE{\(\sigma_{\min}(J^{(k)})>\epsilon_{\sigma}\)}
	\STATE Compute the Jacobian and associated stability quantities \(J^{(k)}\), \(\sigma^{(k)}\), and \(c^{(k)}\) based on Eq.~\eqref{eq:sigma_variation}.
	\STATE Form the local sensitivity coefficients \(\alpha^{(k)}\), \(\beta^{(k)}\), and \(\gamma^{(k)}\) based on Eq. \eqref{eq:direction_basic}, and construct the corresponding reference patterns \(w_P^{(k)}\) and \(w_Q^{(k)}\) based on Eq. \eqref{eq:reference_pattern}.
	\STATE Determine the feasible interval \(\mathcal{B}^{(k)}\).
	\STATE Evaluate \(\Psi^{(k)}(b)\) by solving three allocation subproblems, and obtain \(b_k^{\star}\) based on Eq. \eqref{eq:b_search}.
	\STATE Recover direction \(p_{L,k}^{\star}\), \(g_{P,k}^{\star}\), and \(g_{Q,k}^{\star}\), and set
	\(d^{(k)}=-E_Lp_{L,k}^{\star}+E_Pg_{P,k}^{\star}+E_Qg_{Q,k}^{\star}\).
	\STATE Update power injection \(\rho^{(k+1)}=\rho^{(k)}+\Delta\lambda d^{(k)}\), and solve power flow \(f(x^{(k+1)};\rho^{(k+1)})=0\).
	\STATE Update path-coupled margin \(\eta_{\mathrm{pc}}\leftarrow \eta_{\mathrm{pc}}+\Delta\lambda b_k^{\star}\), and set \(k\leftarrow k+1\).
	\ENDWHILE
	\STATE \textbf{Output:} path-coupled voltage stability margin \(\eta_{\mathrm{pc}}\).
	\end{algorithmic}
\end{algorithm}

\section{Redispatch-Based Margin Improvement and Marginal Stability-Cost Evaluation}
\label{sec_c2_c3}

This section develops the redispatch-based margin improvement method and the associated marginal stability-cost evaluation. It first derives an endpoint-based margin sensitivity that quantifies the first-order effect of generator active- and reactive-power redispatch on the assessed path-coupled margin. This sensitivity is then incorporated into a constrained optimization problem to determine a feasible margin-improving redispatch direction under generator capability limits and active-power balance requirements. Bases on this direction, this section further evaluates the operating cost associated with margin improvement and defines a marginal stability cost as a local price signal for voltage-stability support.

\subsection{Margin-Improving Redispatch Direction}
\label{sub_sec_c2}

The path-coupled assessment provides both the margin value and the collapse point reached from the current operating state. These results can be further used to guide preventive generator redispatch for voltage stability enhancement. Directly deriving the sensitivity of the path-coupled margin \(\eta_{\mathrm{pc}}\) with respect to generator outputs \(u=[P_G^{\top},Q_G^{\top}]^{\top}\) would require differentiating through the entire path-coupled continuation process, since a small redispatch can affect the subsequent direction selections and ultimately the collapse point. To avoid this complex pathwise differentiation, we instead construct an endpoint-based approximation from the assessed margin and collapse point. Specifically, the margin-normalized endpoint direction \(d_{\mathrm{pc}}=(\rho^{\mathrm{c}}-\rho^{(0)})/\eta_{\mathrm{pc}}\) is introduced, where \(\rho^{\mathrm{c}}\) denotes the power injection at the assessed collapse point.

The endpoint approximation allows the effect of a small preventive redispatch to be described directly at the assessed collapse point. A redispatch \(\Delta u\) changes the nodal injections at the current operating point by \(B_u\Delta u\), where \(B_u=[E_P\ E_Q]\). Meanwhile, because \(d_{\mathrm{pc}}\) relates the net endpoint injection change to the path-coupled margin, a first-order margin change \(\Delta\eta_{\mathrm{pc}}\) corresponds to an endpoint injection variation \(d_{\mathrm{pc}}\Delta\eta_{\mathrm{pc}}\). Linearizing the power-flow equation \(f(x;\rho)=0\) around the assessed collapse point \((x^{\mathrm c},\rho^{\mathrm c})\) then gives
\begin{equation}B_u\Delta u
+
d_{\mathrm{pc}}\Delta\eta_{\mathrm{pc}}
-
J^{\mathrm c}\Delta x^{\mathrm c}
=0.\end{equation}
where \(J^{\mathrm c}\) denotes the singular power-flow Jacobian at the assessed collapse point. Its left null vector \(\ell^{\mathrm c}\) yields \((\ell^{\mathrm c})^\top J^{\mathrm c}=0\). We then left-multiply the above perturbation relation by \((\ell^{\mathrm c})^\top\) to eliminate the collapse-state variation \(\Delta x^{\mathrm c}\). This gives the local margin relation \(\Delta\eta_{\mathrm{pc}}\approx g_\eta^\top\Delta u\), where \(g_\eta\) represents the endpoint-based sensitivity of the path-coupled margin to generator active- and reactive-power redispatch:
\begin{equation}
	g_{\eta}
	=
	-
	\frac{B_u^{\top}\ell^{\mathrm{c}}}
	{(\ell^{\mathrm{c}})^{\top}d_{\mathrm{pc}}}.
	\label{eq:margin_gradient}
\end{equation}

The margin-improving redispatch direction is then determined within the feasible generator adjustment space as:
\begin{align}
	\Delta u^\star \in \arg\max_{\Delta u}\quad
	& g_{\eta}^{\top}\Delta u
	-\frac{\kappa}{2}\lVert \Delta u\rVert_2^2
	\nonumber\\
	\text{s.t.}\quad
	& u^{\min}\leq u^{(0)}+\Delta u\leq u^{\max},
	\label{eq:redispatch_direction}\\
	& s_P^{\top}\Delta u=0.
	\nonumber
\end{align}
where \(u^{\min}\) and \(u^{\max}\) collect the generator active- and reactive-power output limits, and \(s_P\) denotes the active-power generation selector such that \(s_P^\top\Delta u=\mathbf{1}^\top\Delta P_G\). The first term maximizes the predicted first-order margin improvement, while the quadratic term regularizes the redispatch magnitude with \(\kappa>0\). The box constraints enforce generator capability limits, and \(s_P^\top\Delta u=0\) maintains the total active-power generation unchanged, preventing the slack generator from implicitly balancing the redispatch. The resulting \(\Delta u^\star\) defines a feasible local redispatch direction for increasing the path-coupled margin. Its actual adjustment depth can be controlled by applying a small step along this direction.

\subsection{Marginal Stability-Cost Evaluation}
\label{sub_sec_c3}

After the margin-improving redispatch direction is obtained, consider a small redispatch step \(u(\alpha)=u^{(0)}+\alpha\Delta u^{\star}\), where \(\alpha\ge0\) is the step size. According to the margin sensitivity in Eq. \eqref{eq:margin_gradient}, the local margin increment along this direction can be approximated by \(\widehat{\Delta\eta}(\alpha)=\alpha g_{\eta}^{\top}\Delta u^{\star}\). The directional margin derivative \(g_{\eta}^{\top}\Delta u^{\star}\) is positive when \(\Delta u^{\star}\) is a nonzero improving direction. Thus, \(\widehat{\Delta\eta}(\alpha)\) gives the approximated improvement value of the voltage stability margin gained by moving the current operating point along \(\Delta u^{\star}\).

The operating cost increment is evaluated from the generator cost function. In this paper, the standard quadratic active-power generation cost is used:
\begin{equation}
	C(u)=
	\sum_{i\in\mathcal{G}}
	\left(
	a_i P_{G,i}^2+b_i P_{G,i}+c_i
	\right),
	\label{eq:generation_cost}
\end{equation}
where $a_i$, $b_i$, and $c_i$ are the cost coefficients of generator $i$. Let $\Delta u^\star=\left[(\Delta P_G^\star)^\top,(\Delta Q_G^\star)^\top\right]^\top$. Since the generation cost is mainly related to active-power generation, the cost increment only takes the cost of active-power generation into account. In that case, the marginal stability cost at the current operating point is given by:
\begin{align}
\mathrm{MSC}
=
\frac{
\left.\frac{dC(u(\alpha))}{d\alpha}\right|_{\alpha=0}
}{
{\left.\frac{d\widehat{\Delta\eta}(\alpha)}{d\alpha}\right|_{\alpha=0}}
}
=
\frac{\sum_{i\in\mathcal{G}}(2a_iP_{G,i}^{(0)}+b_i)\Delta P_{G,i}^\star
}{
g_\eta^\top\Delta u^\star
}
\label{eq:marginal_stability_cost}
\end{align}

The value of MSC measures the local operating cost associated with one additional unit of voltage stability margin. It therefore provides a quantitative price signal for voltage-stability support at the current operating point.

\section{Case Studies}\label{sec_case}

In this section, the proposed PCMA is evaluated on the IEEE 14-bus, IEEE 30-bus, IEEE 39-bus, IEEE 118-bus, and IEEE 300-bus systems. The initial operating point of each system is obtained by AC-OPF calculation with minimum generation cost.

The proposed method is compared with four methods, including two representative voltage stability assessment approaches and two variants of the proposed framework. The first one is the conventional CPF\cite{zimmerman2011matpower}, where the continuation trajectory follows a prescribed loading path obtained by proportionally increasing the base-case load and generation toward a target operating point with doubled injections. The second one is the feasible shortest-distance (FSD) method proposed in \cite{wang2025assessing}, which searches for the locally shortest collapse direction but does not update the generator response trajectory during the continuation process. To further isolate the contributions of adaptive active-power and reactive-power response, PCMA-GR fixes the generator active-power participation ratio throughout the continuation process, while PCMA-PF performs corrective generator response under a fixed power factor. Moverover, redispatch-based margin improvement and marginal stability cost are also investigated to demonstrate how the assessed path-coupled margin can guide preventive generator redispatch and quantify the operating cost of voltage stability enhancement.

\subsection{Path-Coupled Margin Assessment Comparison}

\begin{table}[t]
	\centering
	\caption{Comparison of voltage stability margin assessment methods.}
	\label{tab:margin_comparison}
	\renewcommand{\arraystretch}{1.08}
	\setlength{\tabcolsep}{8pt}
	\begin{tabular}{clcccc}
	\toprule
	System & Method & Margin
	& $Q_{\mathrm{RD}}$ & Time/s & Steps \\
	\midrule
	\multirow{5}{*}{IEEE 14}
	 & CPF     & 3.680 & 3.163 & 1.10 & 68 \\
	 & FSD     & 1.052 & 2.119 & 1.06 & 14 \\
	 & PCMA-GR & 1.014 & 2.095 & 0.54 & 10 \\
	 & PCMA-PF & 1.427 & 2.004 & 0.77 & 15 \\
	 & PCMA    & 1.429 & 1.998 & 0.67 & 14 \\
	\midrule
	\multirow{5}{*}{IEEE 30}
	 & CPF     & 5.858 & 6.295 & 0.72 & 45 \\
	 & FSD     & 1.566 & 1.604 & 0.90 & 11 \\
	 & PCMA-GR & 1.375 & 1.281 & 0.83 & 10 \\
	 & PCMA-PF & 3.334 & 2.619 & 0.86 & 15 \\
	 & PCMA    & 3.573 & 3.589 & 0.62 & 13 \\
	\midrule
	\multirow{5}{*}{IEEE 39}
	 & CPF     &  9.560 & 10.828 &  0.78 &  48 \\
	 & FSD     & 22.394 &  6.598 & 12.37 & 228 \\
	 & PCMA-GR &  7.162 &  8.151 &  1.51 &  24 \\
	 & PCMA-PF &  9.954 & 10.043 &  1.08 &  33 \\
	 & PCMA    & 32.696 & 16.586 &  2.08 & 101 \\
	\midrule
	\multirow{5}{*}{IEEE 118}
	 & CPF     & 4.343 & 1.219 & 4.01 & 160 \\
	 & FSD     & 0.489 & 1.440 & 1.14 &  14 \\
	 & PCMA-GR & 0.462 & 1.457 & 1.31 &  15 \\
	 & PCMA-PF & 1.287 & 2.208 & 1.05 &  17 \\
	 & PCMA    & 4.071 & 2.729 & 3.28 &  63 \\
	\midrule
	\multirow{5}{*}{IEEE 300}
	 & CPF     & 3.110 & 1.010 & 10.24 & 103 \\
	 & FSD     & 0.326 & 0.396 &  1.14 &   7 \\
	 & PCMA-GR & 0.201 & 0.253 &  1.45 &   8 \\
	 & PCMA-PF & 1.409 & 0.393 &  1.91 &  14 \\
	 & PCMA    & 2.535 & 1.593 &  2.10 &  19 \\
	\bottomrule
	\end{tabular}
\end{table}

Table~\ref{tab:margin_comparison} compares the voltage stability margins assessed by the five methods. For each method, the reported margin is defined as the cumulative active load growth from the common initial operating point to the voltage collapse point reached along its continuation trajectory. Among them, CPF generally produces the largest margins because it follows a prescribed loading trajectory determined by proportional load and generation growth without actively searching for a more critical path toward voltage collapse.

The comparison between FSD and PCMA-GR isolates the influence of the loading-direction criterion because both methods retain a fixed generator active-power participation ratio. FSD updates the loading direction according to the maximum voltage-drop criterion, whereas PCMA-GR selects the direction that induces the fastest local decrease in the minimum singular value of the power-flow Jacobian. As shown in Table~\ref{tab:margin_comparison}, PCMA-GR consistently obtains smaller assessed margins than FSD on all five test systems, indicating that the proposed singular-value-based criterion identifies a more critical continuation direction and reaches the voltage-collapse boundary through a shorter cumulative load-growth path.

The three PCMA-family methods are used to evaluate the proposed path-coupled assessment process. PCMA-GR fixes the active-power response direction while allowing reactive-power response to adapt along the continuation trajectory. In contrast, PCMA-PF optimizes the active-power response direction while restricting the reactive-power response through a fixed generator power factor. Compared with these two constrained variants, PCMA jointly updates the active- and reactive-power response directions according to the evolving system state. The coordinated adjustment of active-power balancing and reactive-power support constructs a more effective feasible continuation trajectory, which consequently results in the largest path-coupled margins on all five systems. The improvement is particularly significant for the IEEE 30, IEEE 39, IEEE 118, and IEEE 300 systems, demonstrating the increasing benefit of the complete path-coupled mechanism considering corrective generator response.

Table~\ref{tab:margin_comparison} also reports the terminal reactive-power response magnitude, defined as
\begin{equation}
Q_{\mathrm{RD}}
=
\left\|
Q_G^{\mathrm{c}}-Q_G^{(0)}
\right\|_1,
\end{equation}
where $Q_G^{(0)}$ and $Q_G^{\mathrm{c}}$ denote the vectors of generator reactive-power outputs at the initial operating point and the assessed collapse point, respectively. This metric quantifies the net reactive-power adjustment between the two endpoints of the assessed trajectory. The terminal response magnitude further illustrates the role of adaptive reactive-power support in the path-coupled assessment. Compared with PCMA-PF, PCMA generally mobilizes larger reactive-power adjustments, particularly on the IEEE 39-, 118-, and 300-bus systems, which is accompanied by substantial improvements in the assessed margin. Table~\ref{tab:margin_comparison} further reports the computation time and the number of continuation steps. Since the proposed PCMA repeatedly updates the loading direction and the corrective generator response along the continuation trajectory, it generally requires more steps than its constrained variants. Nevertheless, the additional direction optimization introduces only limited computational overhead. Even for the IEEE 300 system, the complete path-coupled assessment is completed in 1.39~s, indicating that the proposed framework remains computationally efficient for large-scale voltage stability assessment.

\begin{figure}[t]
	\centering
	\includegraphics[width=\columnwidth]{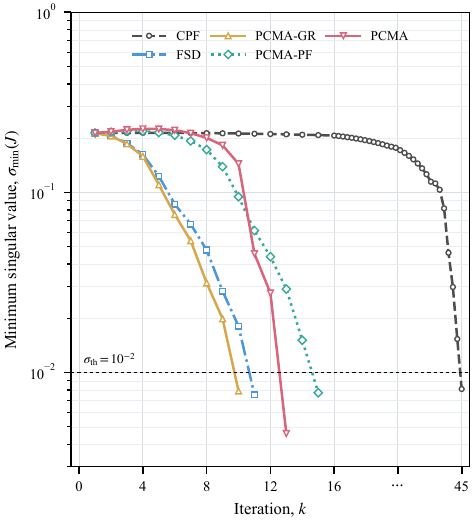}
	\caption{Evolution of the minimum singular value of the power-flow
	Jacobian on the IEEE 30 system versus the continuation iteration.}
	\label{fig:case30_sigma_trace}
\end{figure}

To further investigate the differences among the voltage stability assessment methods, we made more detailed comparison on the IEEE 30-bus system. Fig.~\ref{fig:case30_sigma_trace} illustrates the evolution of the minimum singular value of the power-flow Jacobian during the continuation process of the five methods. CPF gives the longest continuation trajectory among the five methods,. Along this prescribed trajectory, $\sigma_{\min}(J)$ remains relatively high over most of the continuation and drops sharply only near the terminal point. This behavior indicates that the prescribed CPF path is not selected according to the evolving voltage-stability condition and may therefore miss more critical loading directions during the continuation process, leading to a potentially optimistic loading-margin assessment.

Fig.\ref{fig:case30_sigma_trace} further reveals how the remaining four methods approach the Jacobian singularity under different direction-selection and generator response mechanisms. Both FSD and PCMA-GR drive $\sigma_{\min}(J)$ downward from the early continuation stage, indicating that their loading directions actively seek a critical path toward voltage collapse. As discussed above, FSD determines this direction from the steepest decline in bus voltages along the local nose curve, whereas PCMA-GR directly uses the reduction in $\sigma_{\min}(J)$. The more pronounced decrease under PCMA-GR, together with its smaller cumulative load-growth margin in Table~\ref{tab:margin_comparison}, shows that the proposed singular-value-based criterion identifies a trajectory that approaches the Jacobian singularity more critically. A different evolution pattern is observed for PCMA-PF and PCMA. Both methods maintain a relatively high value of $\sigma_{\min}(J)$ during the early continuation stage, indicating that path-coupled corrective response offsets much of the local stability deterioration introduced by adverse load growth. In the later stage, $\sigma_{\min}(J)$ decreases more sharply under PCMA and reaches the termination threshold within fewer iterations than under PCMA-PF. However, as shown in Table~\ref{tab:margin_comparison}, PCMA accumulates a larger active load-growth margin. The combined results indicate that coordinated active- and reactive-power response reshapes the continuation path, allowing the system to accommodate more load growth before the voltage collapse.

\begin{figure}[t]
	\centering
	\includegraphics[width=\columnwidth]{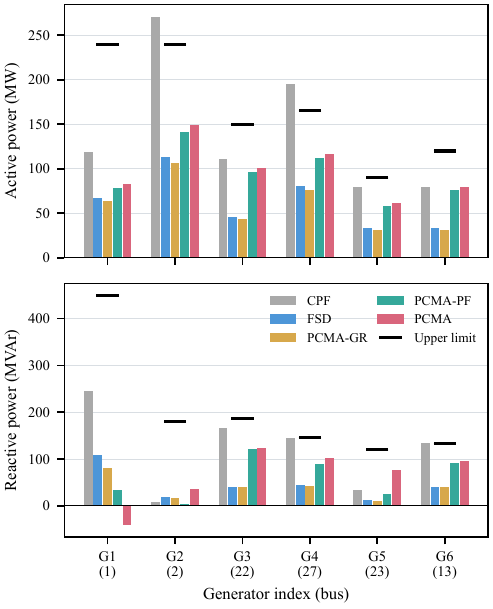}
\caption{Generator active and reactive outputs at the terminal points of
the five assessment methods on the IEEE 30 system.}
	\label{fig:case30_generator_endpoint}
\end{figure}

Fig.\ref{fig:case30_generator_endpoint} compares the generator active- and reactive-power outputs at the terminal points obtained by the five assessment methods. The upper and lower panels show the active- and reactive-power outputs of the six generators, respectively, and the horizontal black markers denote their corresponding upper limits. At the CPF terminal point, the active-power outputs of G2 and G4 reach approximately 271.0MW and 195.2MW, exceeding their upper limits of 240MW and 165MW, respectively. Therefore, the relatively large CPF margin in Table~\ref{tab:margin_comparison} is obtained along a prescribed trajectory that has already left the feasible generator operating region before reaching the assessed collapse point. In contrast, FSD and the three PCMA-based methods maintain all generator outputs within their corresponding limits. Compared with CPF, the proposed PCMA reduces the active-power outputs of G1 and G4 while assigning additional active generation to G2, G3, G5, and G6. On the reactive-power side, G1 absorbs reactive power, whereas the other generators provide different levels of reactive support. These results show that the path-coupled assessment can effectively account for corrective generator response by coordinating active-power balancing and reactive-power support among generators throughout the continuation process.

\subsection{Redispatch-Based Margin Improvement}

\begin{table}
	\centering
	\caption{Redispatch-based margin improvement}
	\label{tab:redispatch_margin_improvement}
	\renewcommand{\arraystretch}{1.2}
	\setlength{\tabcolsep}{5pt}
	\begin{tabular}{@{}lcccccc@{}}
	\toprule
	System & $\eta^0$ & $\eta^{\mathrm{rd}}$ & $\widehat{\Delta\eta}_{\mathrm{sens}}$ & $\displaystyle\frac{\Delta\eta}{\eta^0}$ & $\displaystyle\frac{\Delta\sigma_{\min}}{\sigma_{\min}^0}$ & $E_{\mathrm{rd}}$ \\
	\midrule
	IEEE 14 & 1.4288 & 1.4361 & +0.0075 & +0.51\% & +0.11\% & 0.0210 \\
	IEEE 30 & 3.5731 & 3.5866 & +0.0137 & +0.38\% & +0.66\% & 0.0429 \\
	NE 39 & 32.6959 & 32.7291 & +0.0333 & +0.10\% & +0.20\% & 0.0017 \\
	IEEE 118 & 4.0711 & 4.1706 & +0.1005 & +2.44\% & +0.03\% & 0.0030 \\
	IEEE 300 & 2.5345 & 2.5519 & +0.0132 & +0.69\% & +0.06\% & 0.0008 \\
	\bottomrule
	\end{tabular}
\end{table}

Table~\ref{tab:redispatch_margin_improvement} evaluates the effectiveness of the proposed redispatch direction on the five test systems. After redispatch, the path-coupled margin increases in every case. The sensitivity-predicted increments closely match the recalculated margin gains, confirming that the endpoint sensitivity provides an accurate first-order characterization of the path-coupled margin variation. The minimum singular value at the redispatched operating point also increases on all systems. For example, the IEEE 118-bus system achieves a significant margin increase while exhibiting only a \(0.03\%\) increase in \(\sigma_{\min}\). This indicates that the preventive redispatch improves the path-coupled margin primarily by reshaping the subsequent power-flow trajectory and the collapse point reached along it. Overall, the results demonstrate that the proposed feasible redispatch direction consistently improves voltage stability with limited adjustment effort.

\begin{figure}[!t]
	\centering
	\includegraphics[width=\columnwidth]{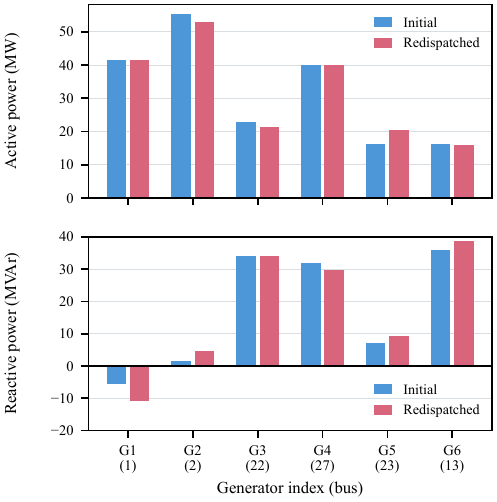}
	\caption{Generator outputs before and after preventive redispatch on the IEEE 30-bus system. }
	\label{fig_case30_redispatch_generation}
\end{figure}

\begin{figure}[!t]
	\centering
	\includegraphics[width=\columnwidth]{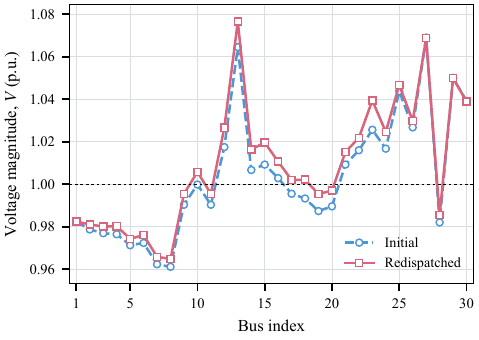}
	\caption{Bus-voltage magnitudes before and after preventive redispatch on
	the IEEE 30-bus system.}
	\label{fig_case30_redispatch_voltage}
\end{figure}

Fig.~\ref{fig_case30_redispatch_generation} further compares the generator output before and after preventive generator redispatch on the IEEE 30-bus system. The active-power adjustment exhibits a clear spatial redistribution, with the outputs of G2, G3, and G4 decreasing, G5 providing the dominant increase, and only minor changes occurring at G1 and G6. This redistribution reflects that adjusting the generation pattern can modify power transfer paths and alleviate critical transmission stresses to improve the path-coupled voltage stability margin. The reactive-power adjustment also exhibits a clear pattern. G2, G5, and G6 increase their reactive-power outputs to provide additional voltage support, whereas G1 absorbs additional reactive power and only minor changes occurring at G3 and G4. Such a coordinated reactive-power redistribution indicates that the margin improvement is obtained by reallocating reactive-power support according to the system voltage-stability characteristics. Fig.~\ref{fig_case30_redispatch_voltage} then shows the voltage magnitude distribution after redispatch. Compared with the original operating point, the voltage magnitudes increase at almost all buses, indicating an overall enhancement of voltage support across the system. The coordinated adjustment of active-power transfer and reactive-power support reshapes the system voltage profile and increases the path-coupled voltage stability margin while preserving the original operating characteristics.

\subsection{Stability-Cost Evaluation}

\begin{figure}[!t]
	\centering
	\includegraphics[width=\columnwidth]{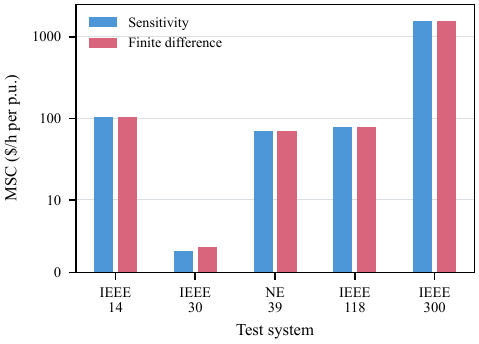}
	\caption{Sensitivity-based and matched finite-difference marginal stability
	costs for the five test systems.}
	\label{fig_stability_cost_systems}
\end{figure}

Fig.~\ref{fig_stability_cost_systems} validates the proposed sensitivity-based marginal stability cost on the five test systems. The proposed sensitivity-based approach directly evaluates the marginal stability cost using the analytical margin sensitivity obtained from the assessed collapse point, whereas the compared finite-difference estimation approximates the same quantity by applying a small redispatch perturbation and recalculating the resulting margin variation. The two approaches show consistent results across all cases, with only small differences observed between the sensitivity-based and finite-difference estimations. This agreement confirms the effectiveness of the endpoint sensitivity in characterizing the variation of the path-coupled margin and enables the marginal cost of voltage-stability enhancement to be evaluated without repeatedly tracing the collapse point.

\begin{figure}[!t]
	\centering
	\includegraphics[width=\columnwidth]{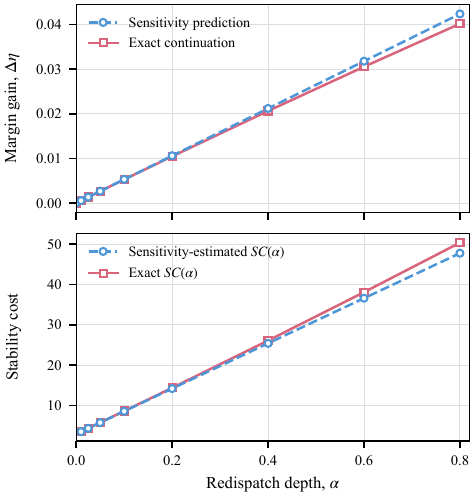}
	\caption{Redispatch-depth-dependent stability-cost evaluation on the IEEE
	30-bus system.}
	\label{fig_case30_redispatch_depth_cost}
\end{figure}

Fig.~\ref{fig_case30_redispatch_depth_cost} further illustrates how the stability improvement and its associated cost vary with the redispatch depth on the IEEE 30-bus system. In this figure, the redispatch depth represents the magnitude of the generator adjustment along the proposed margin-improving direction. The upper subplot compares the margin gain predicted by the analytical sensitivity with the exact gain obtained from continuation calculation, while the lower subplot shows the corresponding stability cost defined as the ratio between the operating-cost increase and the achieved margin improvement. The sensitivity-based prediction closely follows the exact margin gain over the entire redispatch range, showing that the local margin sensitivity remains an accurate approximation along the considered adjustment direction. Meanwhile, the stability cost increases with redispatch depth because the operating-cost increment grows progressively faster due to the quadratic generation-cost characteristics while the margin gain remains approximately linear with redispatch depth. Consequently, a deeper redispatch requires a higher average operating cost for each additional unit of margin gained. The close agreement between the sensitivity-estimated and exact stability-cost curves further shows that the proposed margin sensitivity can accurately characterize this cost variation over the considered redispatch range.

\section{Conclusion}\label{sec_conclusion}
This paper proposes a path-coupled margin assessment method for static voltage stability considering corrective generator response. Compared with conventional continuation-based approaches that rely on predefined load-growth and generation participation paths, the proposed framework jointly determines the adverse load-growth direction and feasible generator response direction during the continuation process, thereby capturing the interaction between loading stress and corrective adjustment actions before voltage collapse. Based on the assessed path-coupled margin, an analytical redispatch direction is further derived for the current operating state by exploiting the saddle-node margin sensitivity, which provides an effective generator redispatch strategy for voltage stability margin improvement. Moreover, a marginal stability cost is introduced to quantify the economic cost associated with increasing voltage stability margin and establish a connection between stability enhancement and operational economy. Future work will investigate the extension of the proposed framework to account for uncertainty-aware voltage stability assessment and coordinated utilization of multiple flexibility resources for preventive stability enhancement.

\footnotesize
\bibliographystyle{IEEEtran}
\bibliography{IEEEabrv,Reference}

\clearpage

\ifCLASSOPTIONcaptionsoff
  \newpage
\fi
\end{document}